\documentclass{jfm}

\usepackage[usenames, dvipsnames]{color}
\usepackage{graphicx}
\usepackage{epstopdf, epsfig}
\usepackage{cleveref}
\usepackage{multirow}

\begin{document}

\newtheorem{lemma}{Lemma}
\newtheorem{corollary}{Corollary}

\shorttitle{Surface seal} 
\shortauthor{A. Kiyama et. al.} 

\title{The water entry surface seal behavior of spheres at intermediate speed regimes}

\author
 {
 Akihito Kiyama\aff{1,2},
 Rafsan Rabbi\aff{1},
 Nathan Speirs\aff{3},
 Jesse Belden\aff{3},
 Yoshiyuki Tagawa\aff{2,4},
 \and 
 Tadd T. Truscott\aff{1,5}
  \corresp{\email{taddtruscott@gmail.com}}
  }

\affiliation
{
\aff{1}
Department of Mechanical and Aerospace Engineering, Utah State University, Logan, UT 84322, USA
\aff{2}
Institute of Global Innovation Research, Tokyo University of Agriculture and Technology, Fuchu, Tokyo 183-8538, Japan

\aff{3}
Naval Undersea Warfare Center Division Newport, 1176 Howell Street, Newport, Rhode Island 02841, USA. 
\aff{4}
Department of Mechanical Systems Engineering, Tokyo University of Agriculture and Technology, Koganei, Tokyo 184-8588, Japan

\aff{5}
Mechanical Engineering Program, Physical Science and Engineering Division, King Abdullah University of Science and Technology (KAUST), Thuwal 23955-6900, Kingdom of Saudi Arabia}
\maketitle

\begin{abstract}  
This research focuses on the water entry of spheres in the  surface seal regime. Herein, surface seal occurs in the wake of a sphere impact with the water surface and is characterized by splash dome over and cavity pull-away between the \emph{low-speed} ($U<30$~m/s) and \emph{high-speed} ($U>165$~m/s) regimes for various sphere diameters $D$. The established empirical scaling laws for pinch-off time $t_p$ hold well for both low and high speeds (i.e., respectively $ t_pUD^{-1}~C_d^{-1/2}=const$ and $ t_pUD^{-1}~C_d^{-1/2}\propto Ca^{-1}$) but the transition deserves refinement.
The intermediate regime in particular scales as $t_pUD^{-1}~C_d^{-1/2}\propto Ca^{-1/2}$, which means that the pressure reduction inside the cavity determines the surface seal dynamics as speeds increase.
We also report on the transition from deep to surface seal, where the seal time is scaled more like a deep seal. The partial pull-away, ripples on the cavity surface, and the secondary pinch-off are also reported.
\end{abstract}

\section{Introduction} 
\label{Intro}
Water entry of a sphere involves complex fluid motion (e.g., \cite{Truscott2014}).
In general, the sphere forms a slender cavity in its wake if the sphere impacts the water surface fast enough \citep{Duez2007}.
The cavities are typically axisymmetric, though they can vary in shape and length.
\cite{Aristoff2009} revealed that these cavities can be categorized into four types: quasi-static, shallow, deep, and surface seals.
\cite{Speirs2019} enlarged the study revealing that the sphere-water wetting angle and surface roughness play a large role in the determination of these cavity types (based on the work of \cite{Duez2007, Zhao2014}).

Herein we focus on surface seal which occurs between \emph{low} and \emph{high} impact speeds where the cavity seals off at the surface from the inward collapse of the splash curtain. The cavity then follows the sphere deeper into the pool.
In particular, we look at the seal time, which is an important parameter for understanding the physics behind the evolution of this cavity type.
Surface seal cavities are often analyzed by their time to seal, which is defined in three ways.
\cite{May1952} was perhaps the first to film three noticeable moments of surface seal, namely the time of ``apparent" surface seal $t_a$ (defined as the moment when any part of the splash starts to fall downwards inside the cavity), the completion of surface seal $t_c$ (defined as the moment when the splash fully collapses on itself, the cavity closes and the overall cavity starts to move downward), and the cavity pull-away $t_p$ (defined as the moment when the cavity separates from the free surface, see section~\ref{results}).
Hereafter, we use $t$ as the general expression of the surface seal time, while we use $t_a, t_c$, and $t_p$ for denoting a specific time as described.

\cite{Lee1997} classified surface seal based on impact speed $U$ into three regimes: \emph{low-speed} ($U<30$~m/s), \emph{high-speed} ($U>165$~m/s), and \emph{very high-speed} ($U>400$~m/s) with most of the past studies focused on the \emph{low-speed} regime.
\cite{Gilbarg1948} performed a \emph{low-speed} surface seal experiment up to $U\approx30$ m/s and found an empirical relationship for the time of surface seal 
\begin{equation}
    t\propto\frac{D}{\rho_gU}, 
    \label{Eq:modelGil}
\end{equation}
where $\rho_g$ is the density of the surrounding gas.
This scaling is estimated using a Bernoulli assumption that the pressure drop inside the cavity is due to the air inflow speed into the cavity that is assumed to be equal to the sphere speed.
Recently, \cite{Eshraghi2020} have estimated the airflow speed by considering the rate of cavity expansion and found that sphere speed  (they tested speeds from 2.0 to 6.0 m/s) is not always a good measure of the airflow speed.
However, the empirical relationship proposed in equation~\ref{Eq:modelGil} has been used to predict the surface seal time for a wide range of parameters.
\cite{May1952} and \cite{Marston2016} experimentally verified that the empirical relationship for the apparent ($t_a$) and pull-away ($t_p$) time of surface seal  $tUD^{-1}\propto\rho_g^{-1}$ holds for various sphere diameters and gas densities.
\cite{Marston2012} found the surface seal time to be  $tUD^{-1}\approx5.75$ under standard atmospheric conditions for spheres heated up to the Leidenfrost temperature and velocities up to 7 m/s.
One exception in the \emph{low-speed} regime is surface seal that occurs near the transition between deep and surface seal entry speeds.
\cite{Eshraghi2020} reported the established scaling law significantly underestimates the surface seal time, which we also discuss in this paper.

\begin{figure}
\centering
\includegraphics[width=0.75\columnwidth]{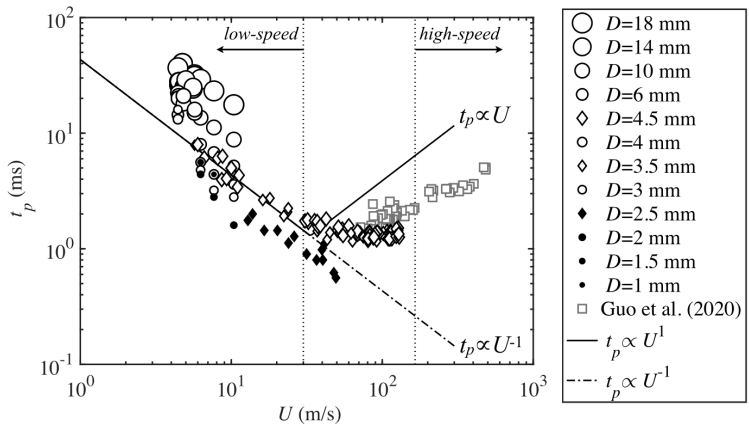}
\caption{A comparison between the cavity pull-away time $t_p$ and sphere impact speed $U$ for various sphere diameters $D$. Circles represent the water entry of sphere, where the experimental conditions are described in section 2. Squares represent the water entry of a cylinder data from \cite{Guo2020}. The solid line shows equation \ref{Eq:modelLee}, where the input pressure $\Delta P$ is switched at $U=34$~m/s and $C_d=0.384$ \citep{Lee1997}. A dashed line shows an extended relationship $t_p\propto U^{-1}$. Vertical dotted lines indicate \emph{low-speed} and \emph{high-speed} regimes.}
\label{fig:intro}
\end{figure}

More importantly, there are only a few studies on the surface seal time for faster speeds ($U>30$~m/s).
\cite{Lee2000} proposed a scaling law that predicts a break-down of equation \ref{Eq:modelGil} at faster sphere speeds.
\cite{Lee2000} considered the characteristic cavity radius $R_{max}$ and the speed of the radial cavity expansion $V_r$ based on the previous consideration \citep{Lee1997} as
\begin{equation}
    t\sim2\frac{R_{max}}{V_r}\propto\frac{\rho_l DU}{\Delta P}~C_d^{1/2},
    \label{Eq:modelLee}
\end{equation}
where $C_d$ is the drag coefficient, see also Supplemental Information.
In brief, the model assumes that the pressure difference  $\Delta P~=P_0-P_c$ is the driving force of the surface seal, where $P_0$ and $P_c$ are respectively the atmospheric pressure and the pressure inside the cavity.
For lower speeds, assuming $\Delta P$ increases as $\Delta P\propto U^2$ based on the Bernoulli equation regarding the airflow, the model predicts $V_r\propto U$ while $R_{max}\propto D \sqrt{(\rho_l/\rho_g)C_d}$ is insensitive to $U$ (see supplemental  information), and thus the surface seal time $t$ decreases as $U$ increases.
This is consistent with the established empirical scaling for \emph{low-speed} surface seal (equation \ref{Eq:modelGil}).
For faster speeds, the model assumes $\Delta P$ stops changing once the pressure inside the cavity reaches the liquid vapour pressure $P_v$, and $V_r$ is not a function of $U$ anymore while $R_{max}$ increases as $U$ increases. 
The model implies that the surface seal time switches from decreasing to an increasing trend at $U = 51$ m/s \citep{Lee2000} but \cite{Abelson1970} showed that this was not the case through pressure measurements where they tried impact speeds of up to 76~m/s and observed a decreasing trend of cavity pressure $P_c$. Further, \cite{Guo2020} also reported that their data at $U$ = 80 to 470 m/s 
did not conform to the theory of \cite{Lee2000,Lee1997}, but did suggest that there was a relationship between the surface seal time and the Cavitation number ($Ca=(P_{0}-P_v)/(\frac{1}{2}\rho_lU^2)$).
We clarify this issue through a visual representation in figure \ref{fig:intro}, where  equation \ref{Eq:modelLee} is plotted over a data set of pull away surface seal time $t_p$ and impact velocity $U$. The fitting is insensitive to $D$ but there is an inflection point where the velocity is assumed to reduce the cavity pressure below the vapor pressure. The data from this study and \cite{Guo2020} are overlaid to show that the theory does a decent job at low velocities but does not account for the transition between low and high speeds. The surface seal time of this transition region (i.e., before the cavity is at vapor pressure) remains worthy of study.

Therefore, this work examines the surface seal at $U>30$~m/s. Our setup allows us to employ the sphere impact speeds up to 128~m/s, which corresponds to the transition regime between \emph{low-speed} and \emph{high-speed} per Lee's definition \citep{Lee1997}, where experimental data are lacking in the literature.
Surface seal time is experimentally evaluated through high-speed imaging observations.
As suggested in equation \ref{Eq:modelLee}, the pressure difference $\Delta P$ plays an important role in determining the surface seal time.
Using $Ca$ to scale the non-dimensional surface seal time makes physical sense since increasing the speed of entry towards total vaporization of the cavity should be considered in the scaling approach.
We discuss the influence of $\Delta P$ on the surface seal time, and show that the scaling laws (equations \ref{Eq:modelGil} \& \ref{Eq:modelLee}) are connected to each other through different interpretation of $\Delta P$ in different speed regimes which yields different $Ca$ based scaling.

This work fills the gap of  experimental understanding up to the \emph{high-speed} surface seal regime and lays the groundwork for future studies in the \emph{very high-speed} regime, where deep seal is predicted to occur rather than surface seal \citep{Lee1997}.

\section{Experimental Methods} 
\begin{figure}
\centering
\includegraphics[width=0.85\columnwidth]{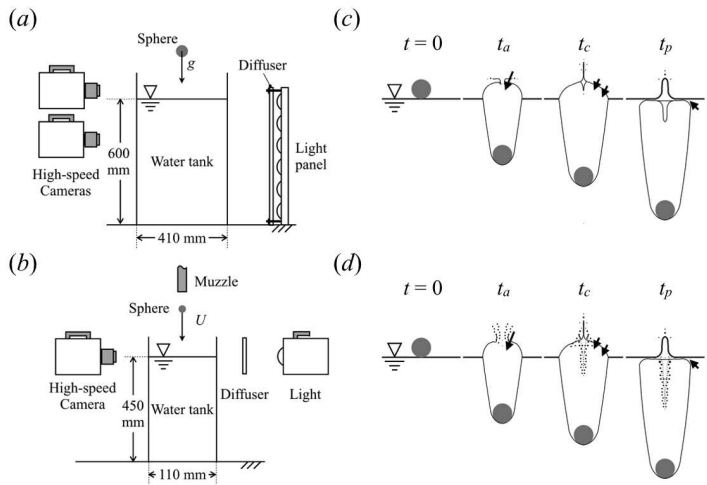}
\caption{The experimental set-ups used; ({\it a}) the free-fall system, and ({\it b}) the air rifle set-up. The impact speed of the sphere was adjusted by changing drop heights for ({\it a}), or the number of air pumps for ({\it b}).
The drops heights for ({\it a}) were set at up to five different levels for each of nine different sphere diameters and the number of pumps for ({\it b}) were set at several different levels for one sphere diameter. Illustrated definitions of time associated with ({\it c}) a slower and ({\it d}) a faster surface seals, $t_a$, $t_c$, and $t_p$ (corresponding to the 4-6th frames in figure \ref{fig:snap1}({\it a-c})). Black arrows indicate the features associated with each definition (section~\ref{results}).}

\label{fig:setup2}
\end{figure}

Two different methods were employed for the varied impact speeds: a free-fall system for the relatively lower impact speeds and an air rifle for the higher speeds as shown in figure \ref{fig:setup2}. The range of parameters are summarized in the Supplemental Information.

The free-fall system is based on that used in \cite{Speirs2019} (figure \ref{fig:setup2}({\it a})).
Stainless steel spheres ($D$ = 1–18 mm, density $\rho_s \approx~$7.8$\times10^3$ kg/m$^3$) were dropped from an electromagnet at a given resting height, resulting in impact speeds from $U=4.43$ to 10.39~m/s.
Spheres are made hydrophobic by applying Glaco Mirror Coat Zero, with an advancing static contact angle of $141^\circ$.
Two high-speed cameras (Photron SA-3) are used to film both splash and cavity dynamics at 2,500 frames per second (fps).

The air rifle setup (figure \ref{fig:setup2}({\it b})) is based on that used in \cite{Kiyama2019}. 
We first use copper-coated spheres (diameter $D\approx$~4.5 mm, density $\rho_s\approx~$6.8$\times10^3$ kg/m$^3$) to shoot from an air rifle (Crossman 760) to the water surface, where the surface seal dynamics are filmed by a high-speed colour camera (Phantom v2510) at 100,000 fps and a resolution of $\approx$~0.1 mm/pix.
The sphere impact speed ranges from $U = 8.13$ to 128~m/s controlled by the number of air pumps.
The spray coating was not applied for faster speed impacts ($U\geq30$~m/s), since it was reported that the surface property does not play a significant role in the cavity dynamics in the high-speed conditions \citep{May1951,Duez2007,Zhao2014}. 
We also capture the overall cavity dynamics with a typical frame rate of 50,000 fps and resolution $\approx$~0.3~mm/pix in a separate experiment.
In addition to the copper-coated spheres ($D~\approx4.5$~mm), we employ two different smaller stainless steel spheres ($D=~2.5$ \& $3.5$~mm,  $\rho_s~\approx~$7.8$\times10^3$ kg/m$^3$).

The image processing including adjustment of the image brightness/contrast was performed through ImageJ. The surface seal times were determined manually based on the definitions described in section~\ref{sec:overall}. The sphere impact speed $U$ for the free-fall set-up was estimated by considering the drop height of the sphere (\cite{Speirs2019}), while that for the air-rifle set-up is calculated by adopting a custom Matlab code that determines velocity from sphere position before impact \citep{epps2010evaluating}. 
For the estimated uncertainty in the image-based measurement, see Supplemental Information.

\section{Results and discussion}
\label{results}
\subsection{Surface seal observations}
\label{sec:overall}
\begin{figure}
\centering
\includegraphics[width=0.75\columnwidth]{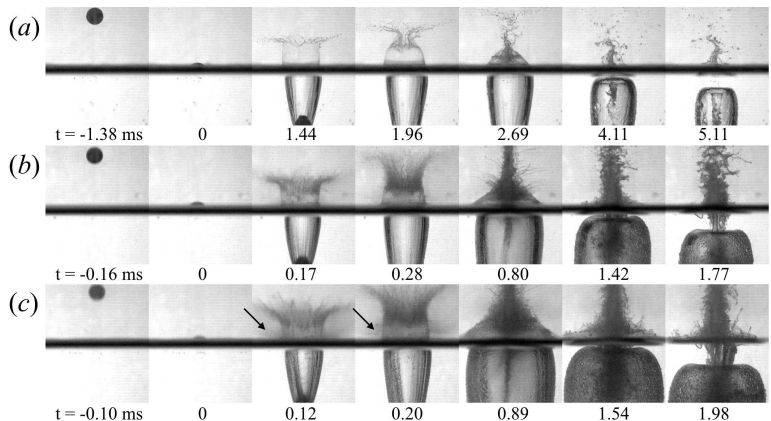}
\caption{Surface seal water entry of three spheres ($D= 4.5$ mm) at velocities: ({\it a}) $U= 9.58$ m/s, ({\it b}) $U= 82.2$ m/s, and ({\it c}) $U= 128$ m/s. $t=0$ is set to the moment when the camera first detects the sphere penetrating the free surface. Arrows point to the residual of the mist-like splash as discussed in section~\ref{results}. Supplemental movies 1-3 correspond to ({\it a-c}).}
\label{fig:snap1}
\end{figure}

We begin by investigating high-speed imagery of surface-seal cavity creation at different speeds. Typical surface seals for three speeds $U=$~9.58, 82.2, and 128 m/s of a $D= 4.5$~mm sphere are shown in figure \ref{fig:snap1} (see also Supplementary videos).
For the slower speed (figure \ref{fig:snap1}({\it a})) a part of the splash sheet starts to fall down when the splash dome has almost closed (1.96 ms).
The splash crown closes on itself at 2.69 ms and then the cavity detaches from the free surface at 4.11 ms, where the downward vertical jet forms inside the cavity.
In contrast, the higher velocities alter the time of surface closure and introduce different features (e.g., \cite{Shi2000}). The higher velocities introduce a mist-like splash and the misty jet inside the cavity (figure \ref{fig:snap1}({\it b})).
This is more pronounced in the fastest case (figure \ref{fig:snap1}({\it c})) where the mist also occurs right above the free surface (indicated by the black arrows).
The conical bubbly regions between the main cavity and free surface (see the last frame of figure \ref{fig:snap1}({\it b}) \& ({\it c})) suggests that the air-fluid mixture is still rushing into the cavity, even if the cavity has pulled away from the free surface.

Classification of surface seal cavity types begins by looking at the times associated with the closure features as sketched in figure~\ref{fig:setup2}({\it c \& d}). 
We define the ``apparent" surface seal $t_a$ as the moment when any part of the splash sheet start to flow downward corresponding to the airflow direction marked by a single arrow.
The ``complete'' seal $t_c$ occurs when the splash crown domes over meeting itself at the central axis.
The dome often starts to move downwards at this moment (double arrows).
The cavity ``pull-away'' $t_p$ occurs when the air cavity descends below the original free surface level.
The 4th, 5th and 6th frames of figure \ref{fig:snap1} respectively correspond to the times $t_a, t_c$ and $t_p$ outlined in figure~\ref{fig:setup2}({\it c \& d}).

We note that in the literature, ``complete" seal (i.e., splash dome-over) is often considered as the indicator of surface seal and ``pull-away" is not \citep{Marston2012,Marston2016,Eshraghi2020}. 
Regarding our images, figure \ref{fig:snap1} indicates that the cavity ``pull-away'' $t_p$ is the easiest to measure; whereas the other two ($t_a$ \& $t_c$) are relatively challenging to measure from these side-view images.
Therefore, we focus on the cavity ``pull-away'' time $t_p$ in this manuscript.
We provide values for other two times ($t_a$ \& $t_c$) in supplemental  information.

\subsection{Time of surface seal for varying sphere speeds and diameters}
\begin{figure}
\centering
\includegraphics[width=0.85\columnwidth]{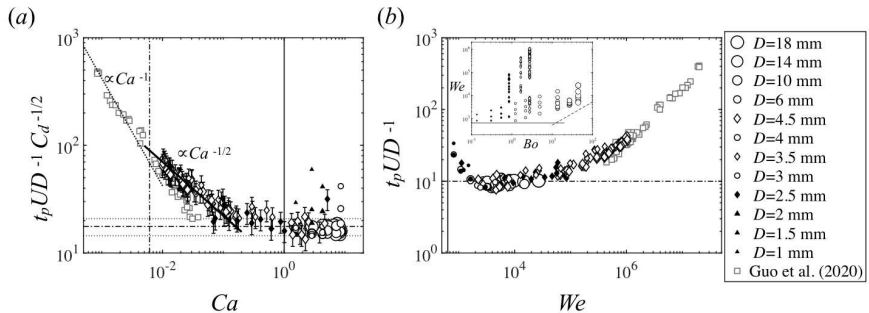}
\caption{The dimensionless time associated with the surface seal  $t_pUD^{-1}C_d^{-1/2}$ as a function of ({\it a}) $Ca$,  while $t_pUD^{-1}$ is plotted against Weber number $We$ in ({\it b}). The size of the markers corresponds to sphere size. Circles represent the data taken by using the set-up shown in figure \ref{fig:setup2}({\it a}) and diamonds the set-up shown in figure \ref{fig:setup2}({\it b}). Sphere diameters smaller than  the capillary length ($l_c \approx 2.7$ mm) are represented by filled markers. Squares represent \cite{Guo2020} data, where we assume $P_0=101.325$~kPa. In ({\it a}), the vertical solid and dot-dashed lines correspond to $Ca=1$ and $Ca=6.1\times10^{-3}$ for a sphere. The horizontal dot-dashed and dotted lines represent the conventional scaling  ($t_pUD^{-1}C_d^{-1/2}=17.6 \pm$3.2) for spheres ($C_d=0.32$).  The error bar reflects variation of $C_d$ values for diamonds (see SI). The inclined dashed and inclined solid lines show the slopes of equations~3.1 \& 3.2 respectively.
The solid vertical line in ({\it b}) indicates the transition threshold from deep seal to surface seal ($We=640$, \cite{Aristoff2009}) and the dashed horizontal line indicates the conventional scaling $t_pUD^{-1}=9.97$. The inset shows a parameter space on Weber $We$ - Bond $Bo$ numbers.}
\label{fig:scaling}
\end{figure}

Our experimental data show that there is a transition regime where the $t_p$ scaling proposed by \cite{Gilbarg1948} (for low speed, equation~\ref{Eq:modelGil}) and \cite{Lee2000} (for high speed, equation~\ref{Eq:modelLee}) does not explain intermediate impact speeds (figure \ref{fig:intro}).
In figure \ref{fig:scaling}, the experimental results of $ t_pUD^{-1}$ are shown as a function of dimensionless parameters $Ca$ and $We$ (figures \ref{fig:scaling}({\it a \& b})).
We also present high-speed experimental data taken from \cite{Guo2020}, who investigated the water entry of a cylinder ($U=80-470$~m/s, $D=6$~mm diameter).
Figure \ref{fig:scaling}({\it a}) indicates that the dimensionless surface seal time $ t_pUD^{-1}C_d^{-1/2}$  for spheres is a constant for $Ca>1$.
Note that few data points for $Ca>1$ do not fit the trend because they have low Bond numbers as discussed later.
A clear separation of data at $Ca\sim O(1)$ suggests that the vaporization of water alters the pull-away mechanism.
Here we consider the scaling law proposed by \cite{Lee1997,Lee2000}, where they considered the cavity pressure $P_c$ as a constant once it reaches the liquid vapour pressure $P_c = P_v$.
Substituting $\Delta P =P_0-P_v$ into equation~\ref{Eq:modelLee} leads to the relationship, 
\begin{eqnarray}
    \frac{t_pU}{D} \propto \frac{\rho_lU^2}{P_0-P_v}~C_d^{1/2},\\\nonumber 
\frac{t_pU}{D} C_d^{-1/2} \propto Ca^{-1},
    \label{eq3.1}
\end{eqnarray}
which scales similar to the data from \cite{Guo2020} (inclined dash-dot line in figure~\ref{fig:scaling}({\it a}),  where the best fit for $Ca<0.01$ was $t_pUD^{-1}C_d^{-1/2}\propto Ca^{-0.89}$, see SI for $C_d$ values used).

Though pressure measurement in our system was not feasible, we found that \cite{Abelson1970} measured the pressure inside the cavity during the water entry of a conical-head projectile (76.2~mm diameter) up to 76.2~m/s.
They reported that the pressure inside the cavity $P_c$ starts to drop when the projectile speed exceeds 10.4~m/s and $P_c$ decreases linearly up to 76.2~m/s.
Substituting $\Delta P\propto U$ \citep{Abelson1970} into equation \ref{Eq:modelLee} leads to another scaling,
\begin{equation}
    \frac{t_pU}{D}~C_d^{-1/2} \propto Ca^{-1/2},
    \label{eq3.2}
\end{equation}
which scales well with  our data (inclined solid line of figure~\ref{fig:scaling}({\it a}),  where the best fit for $0.01<Ca<0.1$ was $t_pUD^{-1}C_d^{-1/2}\propto Ca^{-0.45}$).
Extending the linear relationship until $P_c$ reaches zero pressure, \cite{Abelson1970} predicted the approximate critical velocity of 163~m/s or $Ca\sim6.1\times10^{-3}$ (vertical dot-dashed line) where a transition to \emph{high-speed} occurs (i.e., between the solid line of the \emph{intermediate speed} spheres and the dashed line of \emph{high-speed} cylinders).


Figure \ref{fig:scaling}({\it b}) presents the same data as a function of $We$, indicating that the time of pull-away  $t_pUD^{-1}$ is divided into three regimes. 
First, the scaling law ($t_pUD^{-1}$=const.) holds well for the range $2\times10^3<We<3\times10^4$, where we found the averaged value $t_pUD^{-1}=9.97\pm1.80$ (marked by the horizontal line).
All the data in this regime had sphere speed $U<27$~m/s.
This is consistent with the condition in which no significant cavity pressure drop was recorded by \cite{Abelson1970} (up to 10.4~m/s). 
However, the scaling law  ($t_pUD^{-1}$=const.) does not work well outside of this range.

At $We$ between 640 and 2000,  $t_pUD^{-1}$ decreases as $We$ increases.
This range corresponds to the few outlier cases with low Bond numbers in the $Ca>1$ regime. 
At $We = 640$ surface seal occurs at nearly the same time as deep seal pinch-off resulting in larger  $t_pUD^{-1}$. As $We$ increases the inertial effects become more important than surface tension and the  $t_pUD^{-1}$ values eventually merge to the constant empirical scaling as expected ($We$ between $2\times10^3$ and $3\times10^4$).
The overall trend is consistent with the data reported in \cite{Eshraghi2020}, who studied the surface seal dynamics at slower speeds (up to 6 m/s) and discussed the contribution of the airflow to the surface seal time. They reported the dome over occurred at around $We=1.7\times10^3$.
This slight increase of surface seal time might be qualitatively explained in terms of $\Delta P$ as well. A cavity at $We = 640$ likely experiences both the surface and deep seals at the same time. The airflow could be blocked as the deep seal progresses, resulting in relatively higher pressure inside the cavity and thus smaller $\Delta P$ than that predicted by the Bernoulli approach. The surface seal takes longer to complete as predicted by equation \ref{Eq:modelLee}.

At faster speeds ($We>3\times10^4$), the pull-away time  $t_pUD^{-1}$ increases as $We$ increases.
This regime  largely corresponds to $Ca<1$ in figure \ref{fig:scaling}({\it a}), suggesting that the presence of some water vapor alters the surface seal dynamics.

\subsection{Partial pull-away of the cavity and the ripples on the cavity surface}
\begin{figure}
\centering
\includegraphics[width=0.75\columnwidth]{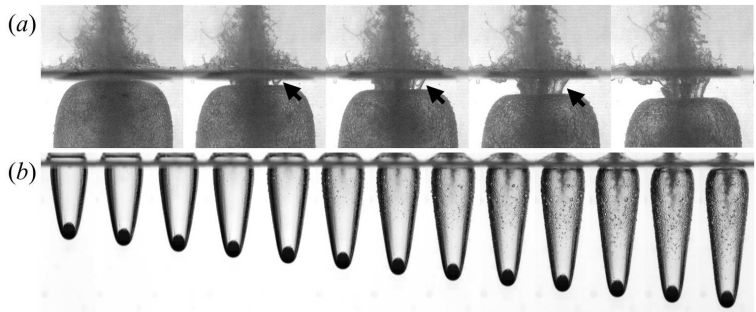}
\caption{({\it a}) Partial pull-away process for $D=4.5$~mm and $U=124$~m/s. The time interval is 0.1~ms. Arrows indicate the air bubbles elongating from the free surface towards the main cavity. ({\it b}) Development of the ripples on the cavity wall for $D=18$~mm and $U=7.67$~m/s. The ripples emanate from droplets from the collapsed splash crown. The time interval is 0.4~ms.}
\label{fig:snap2}
\end{figure}

The partial pull-away is observed when the splash becomes a mist of small droplets as mentioned in the last frames of figure~\ref{fig:snap1}({\it b}) \& ({\it c}).
This phenomenon is triggered because of the emergence of the misty splash sheet upon the fast ejection of a thin splash sheet.
A well-known pull-away is accompanied by the contiguous splash sheet (e.g., figure \ref{fig:snap1}({\it a})), with clear separation between the cavity and the free-surface apparent after pull-away happens.
In contrast, the misty splash is not contiguous, with air filaments connecting the main cavity to the surface upon sheet folding (figure~\ref{fig:snap2}).
The airflow can enter into the cavity through the filaments until they finally collapse on themselves.
 We qualitatively observed that the faster sphere speeds can increase the size or the amount of air bubbles entrapped in the conical filaments connecting the main cavity and the surface  (e.g., figure \ref{fig:snap1}).
The size of the conical portion might also depend on the amount of air bubbles brought from the surface by the low cavity pressure as marked by the arrows in figure \ref{fig:snap2}({\it a}).


The texture seen on the cavity walls (figure \ref{fig:snap2}({\it b})) comes from the impact of small droplets that have broken off of the collapsed splash crown (also seen by \cite{Mansoor2014}).
Each impact forms small, approximately hemispherical cavities that propagate out radially from the impact sites.
These impacts begin near the top of the cavity, where droplets reach first, then move down toward the sphere.
At faster speeds, the crown breaks up into numerous finer droplets that create a rough texture on the cavity wall.

\subsection{Secondary pinch-off}
\begin{figure}
\centering
\includegraphics[width=0.75\columnwidth]{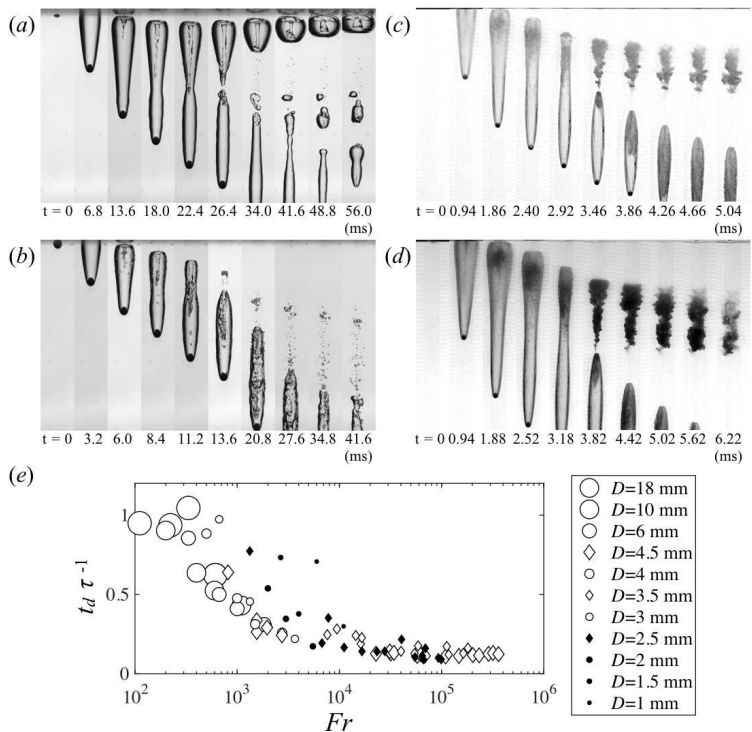}
\caption{Secondary pinch-off of surface seal cavities for ({\it a}) $D=4$ mm, $U=$4.43 m/s, $Fr=5.0\times10^2$, $Ca=8.5$, ({\it b}) $D=4$ mm, $U=$6.26 m/s, $Fr=1.0\times10^3$, $Ca=4.3$, ({\it c}) $D=4.5$ mm, $U=$80.7 m/s, $Fr=1.5\times10^5$, $Ca=2.6\times10^{-2}$, ({\it d}) $D=4.5$ mm, $U=$126 m/s, $Fr=3.6\times10^5$, $Ca=1.1\times10^{-2}$. ({\it e}) The dimensionless pinch-off time of the secondary pinch-off  $t_d~\tau^{-1}$ (corresponds to the 6th frames in ({\it a-d})) as a function of Froude number $Fr$. Supplemental movies 4-7 correspond to ({\it a-d}).} 
\label{fig:second}
\end{figure}

Lastly, we report the dynamics of the secondary pinch-off (i.e., deep seal after surface seal), labeled as $t_d$.
A slower impact leads to surface seal followed by deep seal about half way between the sphere and top of the cavity (see figure \ref{fig:second}({\it a}), $t=26.4$ ms). At higher speeds, the impact introduces different pinch-off dynamics.
The cavity does not collapse at half of its length but near the very top (between frames $t=11.2$ \& $13.6$~ms in figure \ref{fig:second}({\it b})).
The cavity tail shrinks radially and then collides with the vertical jet inside the cavity, perhaps due to the reduced cavity pressure.
The trend holds for much faster speeds as shown in figures~\ref{fig:second}({\it c \& d}).
Although the overall appearance at their pinch-off moments ($t=3.46$~ms in ({\it c}) and 3.82~ms in ({\it d})) are similar in shape to the slower case (figure \ref{fig:second}({\it a})), the pinch-off dynamics are different. The upper portion of the cavity collapses further and forms residual microbubbles.
Since the cavitation number for figures~\ref{fig:second}({\it c \& d}) is much smaller than unity, the residual microbubbles may contain not only the streaming air but also water vapour.

Figure \ref{fig:second}({\it e}) shows $t_d$ measured from the time of sphere impact normalized by $\tau=2.06\sqrt{D/(2g)}$ \citep{Duclaux2007}.
The time  $t_d~\tau^{-1}$ is close to unity for smaller $Fr$ values as expected but decreases as $Fr$ increases (also seen by \cite{Mansoor2014}) until $Fr\sim1.0\times10^4$.
These findings reveal that the reduced cavity pressure makes the secondary pinch-off events depart from typical deep seal behavior as $Fr$ increases.

\section{Conclusion}
Surface seal dynamics between the \emph{low-speed} and \emph{high-speed} regimes ($U=4.43$ to 128~m/s) reveal a strong connection to the decreasing pressure inside the cavity as the water vapour becomes a larger factor in the seal behavior with increasing speed. 
We define the surface seal time by the ``pull-away'' phenomenon $t_p$ and noticed that the dimensionless behavior is more easily scaled by the cavitation number $Ca$. We found that the established semi-empirical scalings hold well for low speeds,  $t_p UD^{-1}C_d^{-1/2}=const$ for $Ca>1$ ($t_p UD^{-1}=9.97\pm1.80$ for $2.0\times10^3<We<3.0\times10^4$), and high speeds, $t_p UD^{-1}C_d^{-1/2} \propto Ca^{-1}$ where $Ca<6.1 \times 10^{-3}$  for cylinders \citep{Guo2020}. 
However, we found the scaling breaks down in the intermediate range and suggest that  $t_pUD^{-1}C_d^{-1/2}\propto Ca^{-1/2}$ at intermediate speeds by assuming $\Delta P\propto U$ as measured by \cite{Abelson1970}. The conventional Bernoulli approach, which describes the overall cavity pressure, is not a good measure to scale the surface seal phenomenon. Instead, as speeds increase beyond $Ca<1$ the cavitation number must be considered, indeed even the recent data of \cite{Guo2020} indicates that this is so.
The cavity dynamics including the partial pull-away (figure \ref{fig:snap2}({\it a})), ripples on the cavity surface (figure \ref{fig:snap2}({\it b})), and the secondary pinch-off (figure \ref{fig:second}) were also discussed.
Increasing sphere speed $U$ also alters the mechanism of the secondary pinch-off from typical deep seal behavior.

\section{Acknowledgments}
A.K. came up with this research while at TUAT. A.K. is currently at USU as JSPS Overseas Research Fellow.
Y.T. acknowledges the financial support from JSPS KAKENHI Grant No. 20H00223.
T.T.T. \& R.R. acknowledge a small portion of the funding from the Office of Naval Research, Navy Undersea Research Program (Grant No. N000141812334). Declaration of Interests: the authors report no conflict of interest.

\bibliography{Ref}
\bibliographystyle{jfm}

\end{document}